\documentclass[12pt]{article}
\usepackage{graphicx} 
\usepackage{cite}

\newcommand{\beq}{\begin{equation}}
\newcommand{\eeq}{\end{equation}}

\begin{document}

\begin{titlepage}

\begin{center}


{\hbox to\hsize {\hfill UW/PT-02-04 }}
{\hbox to\hsize {\hfill  LBNL-49279 }}
\bigskip


\bigskip

\bigskip

\bigskip

{\Large \bf  Yukawa Deflected Gauge Mediation in Four Dimensions}

\bigskip

\bigskip

{\bf Z. Chacko}$^{\bf a,b,c}$,
{\bf E. Katz}$^{\bf d}$, 
and {\bf E. Perazzi}$^{\bf b,e}$ \\

\bigskip

\bigskip

$^{\bf a}${\small \it Department of Physics, University of California,
Berkeley, CA 94720, USA \\
\medskip
$^{\bf b}$ Theoretical Physics Group, Lawrence Berkeley National Laboratory, \\
Berkeley, CA 94720, USA \\ 
\medskip
$^{\bf c}${\rm email}: zchacko@thsrv.lbl.gov \\
$^{\bf e}${\rm email}: EPerazzi@lbl.gov} \\

\smallskip

\bigskip

$^{\bf d}${\small \it Department of
Physics, Box 351560, University of Washington, \\
Seattle, WA 98195, USA\\
\medskip
{\rm email}: amikatz@phys.washington.edu}

\vspace{1cm}

{\bf Abstract}

\end{center}
\noindent

We construct a four dimensional realization of a higher dimensional model, Yukawa
deflected gauge mediation, in which supersymmetry breaking is communicated to the visible
sector through both gauge and Yukawa interactions. The reduction to four dimensions is
achieved by `deconstructing' or `latticizing' the extra dimension. Three sites (gauge
groups) are sufficient to reproduce the spectrum of the higher dimensional model. The
characteristic features of Yukawa deflected gauge mediation, in particular, alignment of
squarks and quarks, and a natural solution to the mu problem, carry over to the
deconstructed version of the model. We comment on the implications of our results for
a solution of the mu problem in the context of deconstructed gaugino mediation.

\end{titlepage}

\renewcommand{\thepage}{\arabic{page}}
\setcounter{page}{1}

\section{Introduction} 

Supersymmetry is perhaps the most attractive solution to the hierarchy
problem.  However, the supertrace theorem implies that SUSY breaking
cannot be communicated at tree level through renormalizable interactions.  
Indeed, tree level breaking would predict superpartners lighter than
observed Standard Model particles.  Consequently, a viable supersymmetric
solution to the hierarchy problem typically includes a ``hidden'' sector
where supersymmetry is spontaneously broken and from where it is
radiatively communicated to the MSSM. The spectrum of superparticle masses
depends crucially on the mechanism through which supersymmetry breaking is
mediated to the MSSM fields.  It thus becomes important to investigate
various forms of mediation of SUSY breaking between the MSSM and the
hidden sector. If the hidden sector and the visible sector are localized
on different branes in a higher dimensional space \cite{RS} then direct
contact operators between the hidden and visible sectors are forbidden by
locality. If the extra dimension is an order of magnitude or so larger in
size than the cutoff of the higher dimensional theory, then the form of the
four dimensional effective theory depends on the light fields in the bulk
and on their couplings to physics on the brane. The various possibilities for
the light fields in the bulk give rise to different patterns of
supersymmetry breaking (for example \cite{RS}, \cite{GLMR}, \cite{KKS},
\cite{CLNP}).

An interesting model, called Yukawa deflected gauge mediation, was 
proposed in \cite{CP}. Here SUSY breaking was
communicated to the MSSM (or visible sector) through gauge mediation
\cite{GMSB}, as well as through mixing between the Higgs and messenger
fields.  As a result, besides the universal gauge mediation soft scalar
masses, there are additional contributions because of direct couplings
between messengers and MSSM fields. These new contributions, which are 
parametrically of the same order as gauge mediation, are
linked to the Yukawa couplings, and thus lead to a mass spectrum with
squark-quark alignment. Hence, the Higgs-messenger mixing of \cite{CP} 
still preserves one of the crucial features of gauge mediation -- namely,
lack of dangerous flavor changing neutral currents (FCNC's). On the other
hand, the model does not suffer from the $\mu$ problem which plagues gauge
mediation.  In the context of the NMSSM, one finds that due to
Higgs-messenger mixing, the singlet field couples more strongly to SUSY
breaking and it is straightforward to obtain the right pattern of
electroweak breaking. 

\begin{figure}
\caption{The three site model}
\center\includegraphics[width=4.5in]{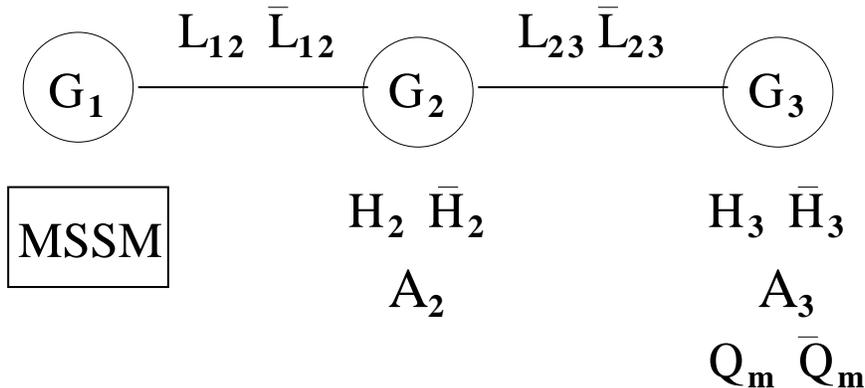}
\end{figure}

The setup of \cite{CP} crucially relied on an extra-dimensional mechanism
to restrict interactions between the MSSM and the messengers. There it was
assumed that the MSSM and the messengers live on seperate branes, while
gauge and Higgs fields propogate in the bulk. In this paper, inspired by
recent ideas of deconstruction (or latticization) \cite{ACG},\cite{HPW} of
higher dimensional dynamics, we propose a entirely four dimensional setup
which realizes the same physics as the higher dimensional setup of
\cite{CP}.

As opposed to genuine extra-dimensional theories, theories with deconstructed extra dimensions
do not have any dimensional coupling, and therefore do not require an UV cutoff.  
Deconstructed extra-dimensions  
have been used earlier to obtain
four dimensional realizations of gaugino mediation \cite{CEGK}, 
\cite{CKSS}. They have also appeared in connection with, among other 
things, new attempts to 
address the hierarchy problem (\cite{ACG2} \cite{CHPW} \cite{CHW}),
models of GUT symmetry breaking (\cite{CKT}, \cite{CMW}), low scale 
unification (\cite{N}, \cite{ACG3}, \cite{CFP}) and fermion localization (\cite{SS}).
Other interesting, more formal results involving deconstruction may 
be found in, for example 
\cite{A},\cite{S},\cite{B},\cite{H},\cite{RS'},\cite{AF},\cite{ACKKM}
\cite{CEKPSS}.

In the next section we present a three link model which is a four
dimensional realization of \cite{CP}.  Here, the link vevs will be much
larger than the messenger scale, and one obtains gauge mediation with the
appropriate Higgs-messenger mixing. Section 3 will address the $\mu$
problem within the context of this model. We conclude in section 4 and
speculate on two possible $\mu$ problem solutions in 
the context of deconstructed gaugino mediation.

\section{A gauge-mediation model with Higgs-messenger mixing}

The model consists of three sites.  Each site indicates a gauge group,
with $G_1 = SM$, $G_2 = SU(5) \times U(1)$, and $G_3 = SU(5) \times U(1)$.  
The field content associated with each site is shown in the figure 1, and
corresponds to a straightforward theory space embedding of the model of
\cite{CP}. The product of $SM \times SU(5) \times SU(5)$ is broken down to
the diagonal standard model by the VEVs of the link fields $L$ and
$\bar{L}$. The additional U(1) factors (which are also broken by the link
fields) and the adjoints $A_2$ and $A_3$ serve only to lift fields which
would otherwise remain massless after symmetry breaking.  As we will
see, having three sites is the minimum required to obtain a light
Higgs field below the link vev scale without fine tuning. 
  
The bifundamental link fields transform as $L_{12}(\bar{5},5,1)$,
$L_{23}(1,\bar{5},5)$, under the three non-abelian gauge factors,
with $\bar{L}_{12}$, $\bar{L}_{23}$ in the
conjugate representations, respectively. Here, for notational simplicity,
we have labeled the transformation of $L_{12}$ under the SM in terms of 
the
transformation of the $\bar{5}$ representation under the SM subgroup of
$SU(5)$.  
The first lattice point hosts the quark and lepton
superfields of the MSSM with the usual standard model charge assignments,
along with the Higgs fields $H'_u$ and $H'_d$.  
The second site hosts  Higgs type field $H_2$ in fundamental representation of $SU(5)$
together with an antifundamental $\bar{H_2}$, and an adjoint $A_2$. 
The third site hosts an Higgs type fundamental $H_3$
and antifundamental $\bar{H_3}$, an adjoint $A_3$, plus the messengers, namely 
two other fundamentals of $SU(5)$, $Q_m$ ($m=1,2$), 
with their conjugates
$\bar{Q}_m$.

The charge assignments of the above fields under the additional $U(1)$ factors
are summarized below.

\begin{equation}
\begin{array}{c|rr}
&U(1)_1 & U(1)_2\\ \hline
L_{12} & -1 & 0 \\
\bar{L}_{12} & +1 & 0 \\
L_{23} & +1 & -1 \\
\bar{L}_{23} & -1 & +1 \\
H_2 & -1 & 0\\
\bar{H}_2 & +1 & 0\\
H_3 & 0 & -1\\
\bar{H}_3 & 0 & +1\\
Q_i & 0 & -1\\
\bar{Q}_i & 0 & +1\\
\end{array} \nonumber
\end{equation}

The SU(2) doublets in the $H_i$'s together with $H_u'$ correspond
to the bulk up type Higgs of \cite{CP} while the doublets in the 
$\bar{H_i}$ together with $H_d'$ correspond to the bulk down type Higgs. 
The quark and lepton superfields localized on the first link correspond to 
having the chiral matter of the MSSM on the brane in \cite{CP}.
The $Q_i$'s and  $\bar{Q}_i$'s
correspond to the messengers localized on the 
hidden brane in the higher dimensional model.

The messengers couple to $H_3$ and  $\bar{H}_3$ through the
superpotential interaction.

\beq 
W_{\rm mix} = X (\lambda_m Q_m \bar{Q}_m + \lambda_d Q_{1}\bar{H}_3 + \lambda_u H_3
\bar{Q}_{2}).
\eeq 

Here $X$ is the spurion with the SUSY violating vev. This is the most
general renormalizable coupling between these fields subject to a discrete
$Z_2$ symmetry under which $X, Q_1$ and $\bar{Q_{2}}$ are odd while
everything else is even, and a discrete $R$ symmetry under which all
fields are odd.  When $X$ picks up a VEV, this interaction will lead to
messenger-Higgs mixing as in the higher dimensional model.

The five dimensional physics of a bulk Higgs field is captured via a site
hopping interaction and link Higgsing potentials:

\begin{eqnarray} 
W_{\rm link} = && \bar{\gamma}_1
H'_d \bar{L}_{12} H_2 + \gamma_2 \bar{H_2}\bar{L}_{23} H_3 + {\gamma}_1 
H'_u
L_{12}\bar{H}_2 + \bar{\gamma}_2 H_2 L_{23} \bar{H}_3  
\cr && + \sum_{i=2,3}
S_i(L_{i-1,i} \bar{L}_{i-1,i} - \alpha_i^2 \Lambda^2) + tr(\bar{L}_{i-1,i} 
A_i L_{i-1,i}).
\end{eqnarray} 

Here, we assume that both singlets $S_i$ couple to a sector
which experiences strong dynamics and generates a term $S_i \alpha_i^2 
\Lambda^2$.  For example, this could be SQCD with
an equal number, $N=2$, of flavors and colors and the couplings
\beq
\frac{S_i}{M_{Pl}^{2}} \epsilon_{mjkl}\epsilon_{\alpha \beta}
\epsilon_{\alpha' \beta'} Q^{\alpha,m} Q^{\beta,j} Q^{\alpha',k}
Q^{\beta',l}. 
\eeq
Here $Q$ represents the quarks of the SU(2) theory, $m,j,k,l$ are flavor
indices and $\alpha, \beta, \alpha'$ and $\beta'$ are color indices.

It is straightforward to verify that the fields $A_i$ in the adjoint
representation of $G_i$ ensure that there will not be unwanted massless
fields, corresponding to additional flat directions upon Higgsing. 

We impose an additional global $U(1)_R$ symmetry and a discrete global $Z_{2 R}$ 
symmetry in order to restrict the possible couplings in the superpotential. 

We summarize in Table~2 the global symmetry transformations
of the various fields.
\begin{equation}
\begin{array}{c|rrr}
&Z_2 & Z_{2 R} & U(1)_R\\ \hline
X &-&-&2\\
S_1 &+&-&2\\
S_2 &+&-&2\\
A_2 &+&-&2\\
A_3 &+&-&2\\
L_{12} &+&-&2/3\\
\bar{L}_{12} &+&-&-2/3\\
L_{23}&+&-&2/3\\
\bar{L}_{23}&+&-&-2/3\\
H'_u &+&-&2/3\\
H'_d &+&-&2/3\\
H_2 &+&-&2\\
\bar{H}_2 &+&-&2/3\\
H_3&+&-&2\\
\bar{H}_3&+&-&-2/3\\
Q_1&-&-&2/3\\
\bar{Q}_1&+&-&-2/3\\
Q_2&+&-&2\\
\bar{Q}_2&-&-&-2\\
{\rm MSSM}&+&-&2/3 
\end{array} \nonumber
\end{equation}
In the above Table ``MSSM'' refers to all the usual MSSM fields other than the Higgses.

The MSSM part of the superpotential consists of the Yukawa couplings of
$H_u'$ and $H_d'$ to the the MSSM quarks and leptons. 
\beq
W_{\rm MSSM} = {\tilde{y}}_{U,ij} {H'}_u q_i {u^c}_j + {\tilde{y}}_{D,ij} {H'}_d
q_i {d^c}_j +
{\tilde{y}}_{L,ij}{H'}_d l_i {e^c}_j 
\eeq
The complete superpotential is $W = W_{\rm mix} + W_{\rm link} + W_{\rm MSSM}$.
{\footnote {The symmetries we have imposed also allow the terms 
$S_2 L_{23} \bar{L}_{23}$ and $tr(L_{23}A_2 \bar{L}_{23})$. Addition of these terms
does not affect our 
conclusions in any way. With the addition of these terms the superpotential $W$
is the most general allowed by the symmetries we have imposed at the renormalizable 
level.}}

The dynamics of this model is as follows.  We will first assume that the
link vevs are much larger than the messenger scale.  Thus, at scale $\sim
\Lambda$, the gauge group $G_1 \times G_2 \times G_3$ is broken to the
diagonal SM gauge group.  The interaction of the link variables with the
adjoint fields, and the D-terms (which follow from the extra $U(1)$
factors) are enough to insure that there are no massless moduli fields
left.  The low energy theory contains only one linear combination of the
doublets in $H_u'$ and $H_3$, which we denote by $H_u$ and one
linear combination of the doublets in $H_d'$ and $\bar{H}_3$,
which we denote by $H_d$. All the triplets decouple. The relevant part of
the superpotential is

\beq 
W = \Lambda(\bar{\gamma}_1 \alpha_1 H_2 H'_d + \gamma_2 \alpha_2 H_3
\bar{H}_2 + {\gamma}_1 \alpha_1 \bar{H}_2 H'_u + \bar{\gamma}_2 \alpha_2 
\bar{H}_3 H_2 )
\eeq
 
This leads to the following expressions for massless doublets:
\begin{eqnarray}
H_u &=& \frac{\gamma_2 \alpha_2 H'_u - \gamma_1 \alpha_1 H_3}{\sqrt{\gamma_1^2
\alpha_1^2 + \gamma_2^2 \alpha_2^2}} \\ \nonumber
H_d &=& \frac{\bar{\gamma}_2 \alpha_2 H'_d - \bar{\gamma}_1 \alpha_1 \bar{H}_3}
{\sqrt{\bar{\gamma}_1^2
\alpha_1^2 + \bar{\gamma}_2^2 \alpha_2^2}}.
\end{eqnarray}

These fields are the analogs of the zero modes of the Higgs fields in the
five dimensional setup of \cite{CP}.  When the supersymmetry breaking
field $X$ acquires an expectation value (at a scale much below $\Lambda$)
the interaction eqn. (1)  mixes $H_u$ and $H_d$ with the doublets in the
messengers so that the physical Higgs doublets and physical mesengers are
linear combinations of $H_u$ and $H_d$ with the doublets in $\bar{Q}_1$
and $Q_2$. The standard model quarks and leptons then have direct Yukawa
couplings to the physical messengers that are proportional to their Yukawa
couplings to the physical Higgs. This is because both sets of Yukawa
couplings arise only through the couplings of the fields $H_u'$ and $H_d'$
to quarks and leptons.  Upon integrating out the messengers, the soft
parameters generated have the Yukawa deflected gauge mediated form of
\cite{CP}. In particular the new contributions to the soft squark masses
respect the flavor structure of the quarks, and do not lead to large
FCNCs.

\section{An NMSSM solution to the $\mu$ problem}

Generating comparable $\mu$ and $\mu B$ terms of weak scale size is an 
important challenge for a SUSY breaking model.  In the context of
gauge mediation, a possible solution is to add an additional singlet
to the MSSM with couplings
\beq
W = \lambda S H_u H_d - \frac{\kappa}{3} S^3.
\eeq

The singlet $S$ is even and odd, respectively, under the $Z_2$ and $Z_{2 R}$ 
symmetries discussed in the previous section, and has charge $2/3$ under the $U(1)_R$
global symmetry.
 
The hope is then that upon SUSY breaking, the singlet will acquire a soft
negative mass squared, and thus a vev on the order of the weak scale.  
The $B \mu$ term then arrises from a SUSY breaking A-term \beq V =
A_{\lambda} S H_u H_d. \eeq However, in gauge mediation this scenario is
difficult to arrange, because $S$ does not couple directly to fields with
tree-level
SUSY breaking splittings, and so obtains only a very small soft mass at
three loops.  The gauge mediated A-term is also not large enough to
produce a $B \mu$ of the right size. This makes it difficult to achieve
the right pattern of electroweak symmetry breaking
\cite{AG},\cite{AFM},\cite{HMZ}. In the limit in which the A-terms are zero,
the NMSSM part of the superpotential has an R symmetry under which all
fields have charge $2/3$.  Since in gauge mediation the A terms
are small, one typically finds that the pseudo-goldstone boson associated with the
spontaneous breaking of the approximate R symmetry is too 
light.

In the higher dimensional theory a way out is offered by Higgs-messenger
mixing.  Here, the singlet has direct Yukawa couplings to the messenger
fields and so naturally gets a soft mass squared at two loops, and A-terms
are generated at the one loop level.  This then leads to a viable
electroweak breaking scenario, without very light goldstone fields
\cite{CP}. It is interesting to see if the same ideas go through in the 
four dimensional case.

In our model, including an NMSSM singlet will generically lead to couplings
of the form

\beq
\label{singlet}
W =  S (\lambda H'_u H'_d + \lambda_3 H_3 \bar{H}_3 + +\lambda_1 H_3 \bar{Q}_1 
+ \lambda_2 Q_2 \bar{H}_3 + \lambda_Q \bar{Q}_1 Q_2) 
- \frac{\kappa}{3} S^3.
\eeq

These couplings mean that below the link vev scale, the singlet will couple to $H_u$
and $H_d$. Further, since $H_u$ and $H_d$ are linear combinations of the physical
higgs and physical messengers, and also because of the direct couplings to the $Q$'s
and $\bar{Q}$'s the singlet gets a soft mass at two loops which is comparable to the
gauge mediated soft masses.  A-terms are
also generated at one loop from these direct messenger matter couplings. It is then
possible to generate the right pattern of symmetry breaking exactly as in the higher
dimensional model.

\section{Conclusions}

In conclusion, we have demonstrated that it is posssible to realize Yukawa
deflected gauge mediation as a four dimensional theory, using the methods
of deconstruction. Three sites are sufficient to reproduce the physics of
the higher dimensional model. All the consequences of Yukawa deflection in
higher dimensions - alignment of squarks and quarks without squark
degeneracy, a natural $\mu$ term mechanism in the context of the NMSSM,
and new decay channels for for the messengers of gauge mediation - carry over
to the four dimensional case.

We end the discussion with a few comments. In our model we have been considering the case
where the link VEVs are larger than the messenger scale. It is worth asking the question
of what happens if the two scales are interchanged so that the messenger scale $<X>$ is
larger than the link scale $\Lambda$. In this limit the model becomes a variant of the
deconstructed gaugino mediation model of \cite{CEGK}, \cite{CKSS} with the Higgs fields
in the bulk. This limit is particularly interesting in the context of the NMSSM model of
section 3. Even in the absence of mixing between messengers and Higgs, the coupling of
the singlet to the Higgs triplets on the third link can generate a negative mass squared
for the singlet radiatively from running between the messenger scale and the link scale
(where the Higgs triplets are integrated out).  It is possible to use this to generate a
VEV for the singlet which can then serve as a $\mu$ term. It may then be possible to
achieve a viable pattern of electroweak symmetry breaking for the deconstructed gaugino
mediation model without sizable fine tuning. The light pseudo-Goldstone associated with
the approximate R symmetry is however a potential problem. 
{\footnote{This approach is similar in
spirit to that of {\cite{DN}} in the context of the NMSSM with gauge mediation. They
obtain a negative mass for the singlet by coupling it to additional vector like matter.}}

If the messenger number symmetry is relaxed to allow messenger-Higgs mixing
interactions the singlet gets a soft mass at two loops at the scale $\Lambda$.
Trilinear soft A-terms between the singlet and the Higgs are also generated at one
loop. The renormalization group evolution to the scale of the link VEVs where the
Higgs triplets are integrated out will result in an additional negative contribution
to the soft mass of the singlet. The theory below the link scale is just the NMSSM
with a gaugino mediated spectrum for the MSSM squarks, sleptons and gauginos but not
for the Higgs fields or the singlet. It is plausible that in large regions of
parameter space the singlet has a negative mass squared, resulting in a sizable VEV
for the singlet. Because the R symmetry is broken at one loop by the singlet-Higgs
trilinear A term, there is no pseudo-Goldstone.  This seems a promising approach for
a solution to the $\mu$ problem of deconstructed gaugino mediation.

\medskip

{\bf Acknowledgements} \\
We would like to thank Kaustubh Agashe, Hsin Chia Cheng, Ann Nelson, Raman Sundrum and
Neal Weiner for discussions at various stages of this work.

\end{document}